\documentclass[doublecol]{epl2} 
\usepackage{amssymb}

\title{The collapse transition on superhydrophobic surfaces}

\author{H. Kusumaatmaja \and M. L. Blow \and A. Dupuis \and J. M. Yeomans}
\shortauthor{H. Kusumaatmaja \etal}

\institute{                    
  The Rudolf Peierls Centre for Theoretical Physics, Oxford University - 1 Keble Road, Oxford OX1 3NP, U.K.\\
}

\abstract{
We investigate the transition between the Cassie-Baxter and Wenzel states of a slowly evaporating, 
micron-scale drop on a superhydrophobic surface. In two dimensions analytical results show that 
there are two collapse mechanisms. For long posts the drop collapses when it is able to overcome 
the free energy barrier presented by the hydrophobic posts. For short posts, as the drop loses 
volume, its curvature increases allowing it to touch the surface below the posts. We emphasise the 
importance of the contact line retreating across the surface as the drop becomes smaller: this 
often preempts the collapse. In a quasi-three dimensional simulation we find similar behaviour, 
with the additional feature that the drop can de-pin from all but the peripheral posts, so that 
its base resembles an inverted bowl.
}

\begin{document}

\maketitle


\section{Introduction}

It is well-known that the hydrophobic nature of a surface is amplified by its roughness \cite{Quere1,Quere3}. This can 
happen in two different ways. When the liquid drop occupies the spaces between the surface projections, and is 
everywhere in contact with the surface, it is said to be in the collapsed or Wenzel state \cite{Wenzel}. The contact angle is
\begin{eqnarray}
& \cos{\theta_{W}} = r \cos{\theta_e} \,  \label{eq1}
\end{eqnarray}
where $r$ is the ratio between the real surface area and its projection onto the horizontal plane and 
$\theta_e$ is the equilibrium contact angle of the flat surface.
On the other hand, if penetration does not occur and the drop remains balanced on the surface projections with 
air beneath it, it is in the suspended or Cassie-Baxter state \cite{Cassie} with contact angle
\begin{eqnarray}
& \cos{\theta_{CB}} = \Phi \cos{\theta_e} - (1-\Phi) \, , \label{eq2} 
\end{eqnarray}
with $\Phi$ the solid fraction of the surface. 
Both states are (local) minimum of the free energy, but there is often a finite energy barrier opposing the  
transition between them. The magnitude of the energy barrier has been shown to depend on both the  
size of the drop and the roughness of the surface \cite{Ishino1,Patankar1}.

The main aim of this paper is to explore the mechanisms by which the drop spontaneously collapses
 \cite{Mchale1,Quere2}. We consider micron-scale drops, sufficiently large 
that we can ignore thermal fluctuations but smaller than the capillary length so that gravity is not important. 
We focus on the limit where the evaporation timescale is much longer than the timescale for drop equilibration 
so that the drop is always in thermodynamic equilibrium. This is normally the physically relevant situation 
for experiments on micron scale drops. The question of how and when collapse occurs is important because, 
even though both states show high values of the contact angle, many of their other physical properties, for 
example, contact angle hysteresis are very different \cite{Kusumaatmaja1}.  

We first consider a drop on a two dimensional, superhydrophobic
surface and present analytic results for how it collapses as its
volume is decreased. We argue that there are two mechanisms for 
collapse. For short posts, as the curvature of the drop increases, it
touches the surface below the posts, thus breaching the free energy
barrier. For longer posts the free energy barrier is removed when the
surface free energy gained by the drop as it collapses wins over
the surface free energy lost by increased contact with the hydrophobic
posts. However, importantly, the collapse transition is usually preempted by
the contact line of the drop retreating across the surface. Therefore
collapse for drops on long posts will normally occur only when the drop
covers a very small number of the posts. 

In three dimensions anaytical calculations are not feasible so we use
numerical simulations to follow the behaviour of the shrinking drop. 
A new feature is that the base of the drop tends to form a bowl-shape, 
where the lines of contact depin and move down all but the outermost posts. 
We further argue that the tendency for the contact line to prefer to 
retreat across the surface than to collapse is even more pronounced in 
three dimensions than in two. A conclusion summarises our results and
compares them to experiments.


\section{Drop collapse in two dimensions: analytical results}

\begin{figure} 
\begin{center}
\includegraphics[scale=0.45,angle=0]{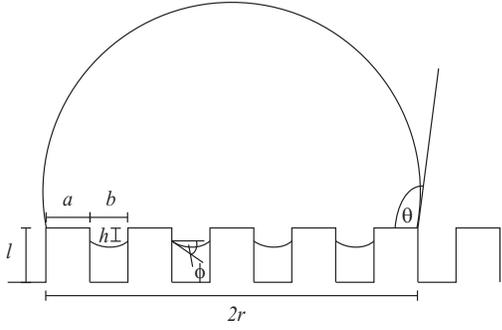}
\caption{Schematic diagram of a suspended drop.} \label{fig1}
\end{center}
\end{figure}

Consider a two dimensional drop suspended on a regular array of hydrophobic posts as shown in Fig. \ref{fig1}. The  
posts have width $a$, spacing $b$ and height $l$, and the substrate material has an intrinsic contact angle  
$\theta_e>90^{\mathrm{o}}$. The drop forms a circular cap with a contact angle $\theta$, cross-sectional area $S$, radius of  
curvature $R$ and base length $2r$.

We consider a drop with contact line that is pinned at the outer edges
of two posts and we first assume that the contact line pinning
persists as the drop spontaneously collapses. Considering the motion 
of a retreating contact line across a superhydrophobic surface (as, 
say, the volume of the drop is slowly decreased) shows that the line is pinned for
$180^{\mathrm{o}}>\theta>\theta_e$   \cite{Kusumaatmaja1}. After we
have explained the possible collapse mechanisms we shall return to a
discussion of   when they are preempted by depinning.

Given pinning, the base radius $r$ is fixed and takes discrete values
\begin{equation}
r = (m+1/2) \, a+m \, b \nonumber
\label{geometry}
\end{equation}
where $2m+1 = 1, 2, 3, \ldots$ is the number of posts beneath the drop. The cross-sectional area of the drop, which is  
constant, can be written 
\begin{equation}
S =  r^2 \, \frac{\theta-\sin{\theta}\cos{\theta}}{\sin^2{\theta}} + 2 \, 
m \, b \, h + \frac{2 \, m \, b^2}{4} \, \frac{\phi-\sin{\phi}\cos{\phi}}{\sin^2{\phi}}.
\, \label{eq3}
\end{equation}
The last term in Eq.~(\ref{eq3}) is due to the curved interface underneath the drop and $\phi = \theta_p -  
90^{\mathrm{o}}$ where $\theta_p$ is the angle this interface makes with the sides of the posts.

Our aim is to investigate when and how the collapse transition occurs. We do this by considering the behaviour of  
the drop free energy as a function of $h$, the distance it penetrates into the substrate (see Fig.~\ref{fig1}).  
The non-constant contributions to the drop free energy $F$ come from three terms. The first two correspond to the  
liquid--gas interfacial free energy above and beneath the surface and the third term is the free energy required by the liquid  
drop to wet the posts to a depth $h$ 
\begin{eqnarray}
& f \equiv F/\gamma = \frac{2\,r\,\theta}{\sin{\theta}} + \frac{2\,m\,b\,\phi}{\sin{\phi}} - 
4\,m\,h\,\cos{\theta_e} \, \label{eq5} 
\end{eqnarray}
where $\gamma$ is the liquid--gas interfacial tension.

We now consider the variation of the free energy with $h$. The drop will  
start to penetrate the posts if $\frac{df}{dh}<0$ at $h=0$, or equivalently $\frac{df}{d\theta}>0$, since  
$\frac{dh}{d\theta}<0$. Using the constraint of constant area to eliminate $dh$ gives
\begin{eqnarray}
df &=& \frac{2r\,(\sin{\theta} - \theta\,\cos{\theta})}{\sin^3{\theta}} \, (\sin{\theta} 
+ \frac{2r}{b}\cos{\theta_e}) \, d\theta + \nonumber \\
   &&  \frac{2mb\,(\sin{\phi} - \phi\,\cos{\phi})}{\sin^3{\phi}} \, (\sin{\phi} + 
\cos{\theta_e}) \, d\phi \, . \label{eq13}
\end{eqnarray}

Consider first $d\phi=0$. Since $2\,r\,(\sin{\theta} - \theta\,\cos{\theta})/\sin^3{\theta} > 0$, 
the condition for the drop to start collapsing is 
\begin{equation}
\sin{\theta} + \frac{2r}{b}\cos{\theta_e} > 0 |_{h=0} \, . \label{eq6}
\end{equation}
The corresponding critical drop radius of curvature and contact angle are \cite{Barrat}
\begin{eqnarray}
& R_c = - \frac{b}{2\cos{\theta_e}} \, , \label{eq23} \\
& \sin{\theta_c} = - \frac{2r}{b}\cos{\theta_e} \, . \label{eq24}
\end{eqnarray}
$\theta$ gets smaller and $\sin{\theta}$ gets larger as the drop penetrates the posts. As a result, once Eq.  
(\ref{eq6}) is satisfied it will always be satisfied and once the drop has started  
to move it collapses fully, into the Wenzel state. 

The drop will be in equilibrium at $h=0$ on the threshold of the collapse transition. Therefore we may combine  
Eq. (\ref{eq24}) and the Laplace pressure condition to show that $\theta_p$ = $\theta_e$, or $\phi = \theta_e - 90^{\mathrm{o}}$ as  
expected from the Gibbs' criterion \cite{Gibbs}. Hence, from Eq. (\ref{eq13}) the free energy is at an extremum  
with respect to changes in $\phi$. Calculating the second derivative confirms that this is a minimum and hence  
that the assumption $d \phi =0$ is appropriate.

\begin{figure} 
\begin{center}
\includegraphics[scale=0.7,angle=0]{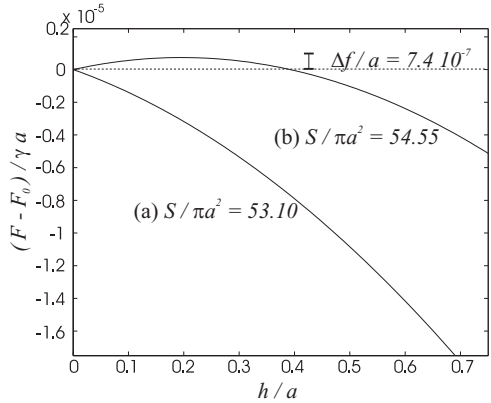}
\caption{
Normalised drop free energy against penetration depth when (a) the collapse transition occurs and (b) there is a free  
energy barrier between Cassie-Baxter and Wenzel states. $F_0$ is the drop free energy in the Cassie-Baxter state,  
$\gamma$ is the liquid--gas surface tension and $S$ is the drop area. $m=3$, $b/a=1.5$, $\theta_e =  
95^{\mathrm{o}}$ and $\theta |_{h=0} = 111^{\mathrm{o}}$ and $112^{\mathrm{o}}$ for (a) and (b) respectively.}  
\label{fig2}
\end{center}
\end{figure}

Typical plots of the free energy of a drop against $h$, the distance it penetrates into the substrate are shown in  
Fig. \ref{fig2}, where for simplicity we have neglected the corrections due to the curvature of the interfaces in  
the grooves. In Fig. \ref{fig2}(a), where we have used $m=3$, $b/a=1.5$, $\theta_e = 95^{\mathrm{o}}$, and $\theta  
|_{h=0} = 111^{\mathrm{o}} < \theta_c = 111.6^{\mathrm{o}}$ the free energy is a smoothly decreasing function of  
$h$ and the drop will collapse. In Fig. \ref{fig2}(b) on the other hand, for $\theta|_{h=0} =  
112^{\mathrm{o}} > \theta_c = 111.6^{\mathrm{o}}$, there is a free energy barrier and therefore no collapse. The  
peak of the free energy barrier occurs at $\theta = \theta_c$ and the magnitude of the barrier is  
\begin{eqnarray}
\Delta{f} &=& \frac{2\,r\,\theta_c}{\sin{\theta_c}} + \frac{2\,r^2\,\cos{\theta_e}}{b}\,\frac{\theta_c-\sin{\theta_c}\cos{\theta_c}}{\sin^2{\theta_c}} \label{eq16} \\
&-& \left[\frac{2\,r\,\theta}{\sin{\theta}} + \frac{2\,r^2\,\cos{\theta_e}}{b}\,\frac{\theta-\sin{\theta}\cos{\theta}}{\sin^2{\theta}}\right]_{\theta\equiv\theta |_{h=0}} \, . \nonumber
\end{eqnarray}

We have argued that, for $R < R_c$, there is no free energy barrier to drops penetrating hydrophobic posts. The  
critical radius depends on the post width $a$, the post separation $b$, the base radius $r$, and the  
equilibrium contact angle $\theta_e$. It does not, however, depend on the post height $l$. There is, however,  
another route to drop collapse \cite{Quere2}, which will pre-empt this mechanism for shallow posts.

Prior to collapse the liquid drop has not penetrated the posts, the system is in mechanical equilibrium, and the  
Laplace pressure is the same everywhere. Thus the liquid--gas interface between the posts bows out with a radius  
of curvature equal to that of the circular cap $R$. The centre of the curved interface reaches a distance $d$ into  
the posts:
\begin{equation}
d  = R \, (1- \cos{\phi}) \simeq \frac{b^2}{8R} \label{eq18}
\end{equation} 
for small $\phi$. As $R$ gets smaller, $d$ increases. When $d=l$ the liquid--gas interface touches the base  
surface initiating the transition between the Cassie-Baxter and Wenzel states. At this point there is a  
considerable free energy release because the drop is replacing two interfaces (liquid--gas and gas--solid) with a  
single liquid--solid interface. Consequently this transition is irreversible and for the opposite transition to  
occur (Wenzel to Cassie-Baxter) an external force is needed to overcome the free energy barrier. For  
this mechanism to be possible it is apparent from simple geometry that $d<b/2$.

\begin{figure} 
\begin{center}
\includegraphics[scale=0.65,angle=0]{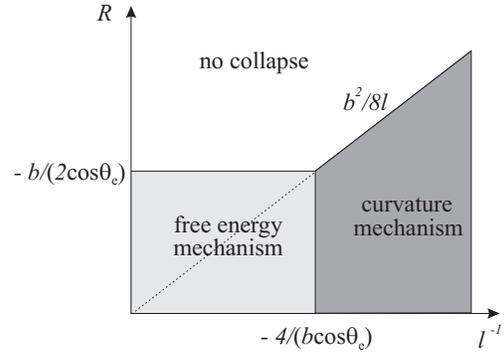}
\caption{The crossover between the two different drop collapse mechanisms in two dimensions.} \label{fig7}
\end{center}
\end{figure}

Regions of parameter space where there is (i) collapse due to the contact line sliding down the posts, (ii)  
collapse due to the centre of the interface touching the base surface, (iii) no collapse are distiguished in Fig.  
\ref{fig7}. The crossover between regions (i) and (ii) occurs when
\begin{equation}
\cos{\theta_e} < -\frac{4\,l}{b}. \label{eq17}
\end{equation}
It is interesting to note that the crossover  
point between the two regimes (Eq. \ref{eq17}) will slide to larger $l$ as the posts are made more hydrophobic. 

We now revisit the assumption that the contact line is pinned at the outer edges of the posts. The (theoretical)  
advancing contact angle is $180^\mathrm{o}$ \cite{Kusumaatmaja1} and therefore the line will not move outwards. 
The receding angle in the quasi-static limit is $\theta_e$ \cite{Kusumaatmaja1} and therefore it will not jump inwards if 
$\theta_c > \theta_e$ or, equivalently, $\sin^2{\theta_c} < 1 - \cos^2{\theta_e}$.
Using Eqs.~(\ref{geometry}) and (\ref{eq24}) this is equivalent to
\begin{equation}
\cos^2{\theta_e} < \left[ 4\left(\frac{m+1/2}{b/a}+m\right)^2+1 \right]^{-1}. \label{eq15}
\end{equation}

\begin{figure*} 
\begin{center}
\includegraphics[scale=1.0,angle=0]{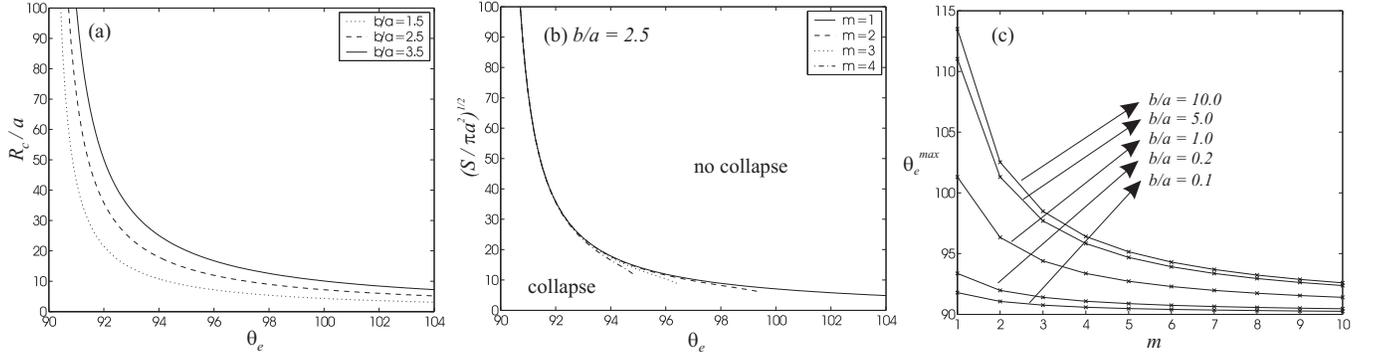}
\caption{(a) Critical drop radius of curvature and (b) area (presented as $\sqrt{S/\pi a^2}$) for collapse as a function of the equilibrium contact angle  
$\theta_e$ for different substrate geometries and number of posts beneath the drop. (c) The maximum value of $\theta_e$ for 
which a spontaneous drop collapse can occur for different b/a and m.} \label{fig4}
\end{center}
\end{figure*}

Fig \ref{fig4}(c) shows the maximum value of $\theta_e$ at which collapse will occur
for different $m$ and $b/a$. For small $b/a$ collapse is
strongly suppressed and only occurs for tiny drops on
slightly hydrophobic surfaces. Even for $b/a >> 1$ the tendency to
depin is strong and the collapse occurs for small value of $m$ unless 
$\theta_e$ is close to $90^{\mathrm{o}}$. By setting $r=b$ in
Eq. (\ref{eq6}), we conclude that the transition will never occur 
spontaneously for $\theta_e > 120^{\mathrm{o}}$. 

In Fig. \ref{fig4}(a) and (b), we plot the critical radius of
curvature at which a transition occurs $R_c/a$ and the corresponding
drop area (which we present as $\sqrt{S/\pi a^2}$) as a function of
$\theta_e$, $m$ and $b/a$. As expected the critical radius of curvature does not
depend on $m$; it is a function of $b/a$ and $\theta_e$ only. The
critical base area of the drop does, however, depend on $m$ and is smaller for
larger values of $m$ and  $a/b$.  The curves for increasing $m$
terminate at decreasing values of $\theta_e$ corresponding to the
contact line receding inwards before the drop is able to penetrate the
posts.

\section{Simulations of drop collapse}

We now describe the details of a numerical model which will allow us to explore the collapse transition in
both two and three dimensions. We describe the equilibrium properties of the drop by a continuum free energy \cite{Briant1}
\begin{equation} 
\Psi = \int_V (\psi_b(n)+\frac{\kappa}{2} (\partial_{\alpha}n)^2) dV + \int_S \psi_s(n_s) dS . \label{eq103}
\end{equation}
$\psi_b(n)$ is a bulk free energy term which we take to be \cite{Briant1}
\begin{equation}
\psi_b (n) = p_c (\nu_n+1)^2 (\nu_n^2-2\nu_n+3-2\beta\tau_w) \, ,
\end{equation}
where $\nu_n = {(n-n_c)}/{n_c}$, $\tau_w = {(T_c-T)}/{T_c}$ and $n$, $n_c$, $T$, $T_c$ and $p_c$ are the local  
density, critical density, local temperature, critical temperature and critical pressure of the fluid  
respectively. This choice of free energy leads to two coexisting bulk phases of density  
$n_c(1\pm\sqrt{\beta\tau_w})$, which represent the liquid drop and surrounding gas respectively. Varying $\beta$  
has the effects of varying the densities, surface tension, and interface width; we typically choose $\beta = 0.1$. 
\begin{figure} 
\begin{center}
\includegraphics[scale=0.7,angle=0]{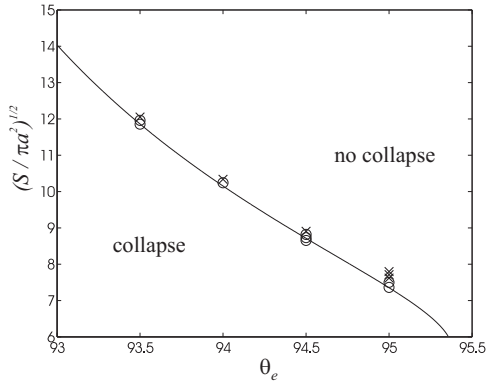}
\caption{
Critical drop area (presented as $\sqrt{S/\pi a^2}$) as a function of contact angle: comparison between the 
two--dimensional theory and simulations for $m=3$, $a = 8$, $b = 12$. The circles and crosses represent drops which collapse and 
remain suspended respectively. The solid line is the theoretical prediction.} \label{fig8}
\end{center}
\end{figure}

The second term in Eq. (\ref{eq103}) models the free energy associated with any interfaces in the system. $\kappa$  
is related to the liquid--gas surface tension and interface width via $\sigma_{lg} = {(4\sqrt{2\kappa p_c}  
(\beta\tau_w)^{3/2} n_c)}/3$ and $\xi = (\kappa n_c^2/4\beta\tau_w p_c)^{1/2}$ \cite{Briant1}. 
We use $\kappa = 0.0018$, $p_c = 1/8$, $\tau_w = 0.3$, and $n_c = 3.5$.

The last term in Eq. (\ref{eq103}) describes the interactions between the fluid and the solid surface. Following  
Cahn \cite{Cahn} the surface energy density is taken to be $\psi_s (n) = -\lambda \, n_s$, where $n_s$ is the value  
of the fluid density at the surface. The strength of interaction, and hence the local equilibrium contact angle,  
is parameterised by the variable $\lambda$. Minimising the free energy (\ref{eq3}) leads to a boundary condition 
at the surface, $\partial_{\perp}n = -\lambda/\kappa$,
and a relation between $\lambda$ and the equilibrium contact angle $\theta_e$ \cite{Briant1}
\begin{equation}
\lambda = 2\beta\tau_w\sqrt{2p_c\kappa} \,\,  
\mathrm{sign}(\frac{\pi}{2}-\theta_e)\sqrt{\cos{\frac{\alpha}{3}}(1-\cos{\frac{\alpha}{3}})} \, , \label{eq106}
\end{equation}
where $\alpha=\cos^{-1}{(\sin^2{\theta_e})}$ and the function sign returns the sign of its argument. Similar  
boundary conditions can be used for surfaces that are not flat: a way to treat the corners and ridges needed to  
model superhydrophobic surfaces is described in \cite{Dupuis2}.

The equations of motion of the drop are the continuity and the Navier-Stokes equations
\begin{eqnarray}
&\partial_{t}n+\partial_{\alpha}(nu_{\alpha})=0 \, , \label{eq104}\\
&\partial_{t}(nu_{\alpha})+\partial_{\beta}(nu_{\alpha}u_{\beta}) = - \partial_{\beta}P_{\alpha\beta} + \nonumber \\
&\nu \partial_{\beta}[n(\partial_{\beta}u_{\alpha} + \partial_{\alpha}u_{\beta} + \delta_{\alpha\beta}  
\partial_{\gamma} u_{\gamma}) ]  \label{eq105}
\end{eqnarray}
where $\mathbf{u}$, $\mathbf{P}$, and $\nu$ are the local velocity, pressure tensor, and kinematic  
viscosity respectively. The thermodynamic properties of the drop appear in the equations of  
motion through the pressure tensor $\mathbf{P}$ which can be calculated from the free energy  
\cite{Briant1,Dupuis2}
\begin{eqnarray} 
&P_{\alpha\beta} = (p_{\mathrm{b}}-\frac{\kappa}{2} (\partial_{\alpha}n)^2 - \kappa n  
\partial_{\gamma\gamma}n)\delta_{\alpha\beta} + \kappa (\partial_{\alpha}n)(\partial_{\beta}n),  \nonumber \\ 
&p_{\mathrm{b}} = p_c (\nu_n+1)^2 (3\nu_n^2-2\nu_n+1-2\beta\tau_w) .
\end{eqnarray}
When the drop is at rest $\partial_{\alpha}P_{\alpha\beta} = 0$ and the 
free energy (\ref{eq103}) is minimised. As we are considering the quasi--static problem when the drop is in  
equilibrium until the point of collapse details of its dynamics should not affect the results. However we choose  
to implement physical equations of motion as this helps the drop to reach equilibrium quickly as its volume is  
decreased and for comparison to possible work on non-equilibrium collapse. 

We use a lattice Boltzmann algorithm to solve Eqs. (\ref{eq104}) and (\ref{eq105}). No-slip boundary conditions on  
the velocity are imposed on the surfaces adjacent to and opposite the drop and periodic boundary conditions are  
used in the two perpendicular directions. Details of the lattice Boltzmann approach and of its application to drop  
dynamics are given in \cite{Briant1,Dupuis2,Succi1,Kwok1,Sbragaglia}. 

To implement evaporation we need to slowly decrease the drop volume. To do this we vary the liquid  
density by $- 0.1\%$ every $2\times 10^5$ time steps to ensure that the evaporation timescale is well separated 
from the drop equilibration timescale. This in turn affects the drop volume as the system relaxes back to its coexisting  
equilibrium densities.
 
Results for two dimensions are compared to the analytic solution in Fig. \ref{fig8}, where we have used 
$a=8$, $b=12$, and $\theta_e = 95^{\mathrm{o}}$.  The critical drop area at which the collapse transition occurs 
is close to the theoretical value but critical radii obtained from simulations are typically too large by $\sim 2$ 
lattice spacings. This is because the liquid--gas interface is diffuse ($\sim 3-4$ lattice spacings). We checked 
that, as expected, $R_c$ is independent of the post height and that $h$ is the same everywhere underneath the drop.

\section{Drop collapse in three dimensions: numerical results}

Analytic calculations in three dimensions are, in general, not
possible for several reasons.  Firstly, the drop shape is not a
spherical cap but is influenced by the underlying topological
patterning. Secondly, the shape  of   the liquid--gas interface
spanning the posts is complicated. Thirdly, $h$, the distance the
drop penetrates the   substrate, is not neccesarily the same
everywhere. Therefore we need to use the numerical approach presented
in the last section to explore collapse. We consider a square array of posts of
widths $a=3$ and spacing $b=9$. We present results for both spherical
drops and `cylindrical'  drops which demonstrate the relevant physics
but are less demanding in computer time.
\begin{figure} 
\begin{center}
\includegraphics[scale=0.675,angle=0]{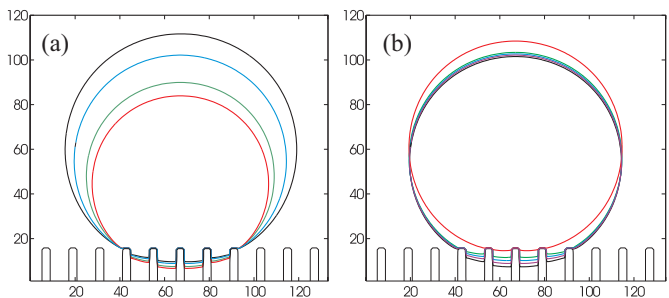}
\caption{ (Color online) Equilibrium drop configurations. (a) $\theta_e = 95^{\mathrm{o}}$ and varying volume. 
(b) Fixed volume and contact angles, $\theta=93^{\mathrm{o}}$ (black), $95^{\mathrm{o}}$ (purple), $97^{\mathrm{o}}$ 
(blue), $100^{\mathrm{o}}$ (green), and $110^{\mathrm{o}}$ (red).} \label{fig6}
\end{center}
\end{figure}

A new feature in three dimensions is that for a spherical (or
cylindrical) drop on a square array of posts the base of the drop can
form a bowl-shape where the lines of contact with the top of all but 
the peripheral posts depin and move down the posts leaving the drop 
suspended by just its outer rim. This was seen in simulations for drops 
of both cylindrical and spherical symmetry, and has recently been 
reported experimentally \cite{Moulinet1}. The depinning occurs to reduce 
the distortion of the interface from spherical. Depinning is
favoured for smaller drops and for contact angles close to
$90^{\mathrm{o}}$. 
\begin{figure}[t]
\begin{center}
\includegraphics[scale=0.9,angle=0]{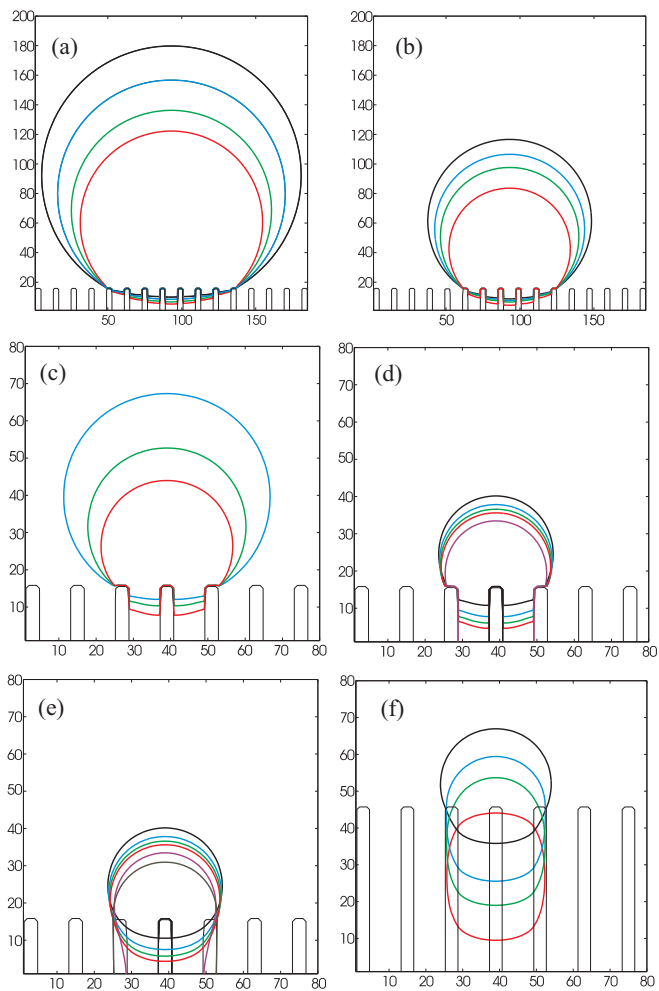}
\caption{
(Color online) Evolution with time of a cylindrical drop on a square array of posts
of width $a=3$, spacing $b=9$ and height $l=15$. (a--c) Time evolution
before collapse showing depinning of the receding contact line (note
the scale change between (b) and (c)). (d--f) Motion of the collapsing drop:
(d) cross sections in the plane bisecting the posts. (e) same times as
(d), but in the plane bisecting the gap between the posts. (f) cross
sections in the plane bisecting the gap, but with $l=45$ to enable the
collapse to be followed to later times. 
} \label{figQuasi2D}
\end{center}
\end{figure}

This is apparent in Fig. \ref{fig6}(a) which shows the equilibrium profile of the drop as its
volume is varied with $\theta_e = 95^{\mathrm{o}}$.  As expected the
drop penetrates further into the posts as the radius is decreased
(corresponding to increasing  curvature). Fig. \ref{fig6}(b) shows
cross sections of the final states of 5 drops with volumes $V
\simeq 4.5 \times 10^{5}$ (in the units of lattice spacing) equal to within $\sim 1\%$ but varying equilibrium
contact angles in the range $\theta_e = 93^{\mathrm{o}}$ to
$110^{\mathrm{o}}$. Although the drop penetrates deeper into the posts
as the intrinsic  contact angle approaches $90^{\mathrm{o}}$ there is
no collapse (these values would give a collapse transition in two
dimensions). Instead, for $\theta_e=110^{\mathrm{o}}$, the contact line
depins and moves to cover 9 rather than 21 posts. Note that this jump 
also corresponds to a transition to the state when the drop is suspended 
on all the posts beneath it, not just those around its rim.

To explore the depinning further, and to try to find a collapse
transition, we turned to the geometry of a cylindrical drop on a square
array of posts. This preserves the physics whilst allowing us to
exploit the quasi-two-dimensional geometry to run larger
simulations. Results are shown in Fig. \ref{figQuasi2D} for 
$\theta_e=93^{\mathrm{o}}$. Successive frames show how the
drop profile evolves as its volume is quasi-statically decreased (note
that they are drawn on different scales). Initially, the contact line 
is pinned at the edges of the posts and the drop penetrates further 
beneath the posts as the radius is decreased. However, as the drop continues 
to decrease in size, the drop contact angle reaches the receding angle
and the contact line depins. As it depins we observe that the penetration into the posts
decreases (because the drop is approximately spherical and the base area is
reduced), thus moving the system away from the point where either a
curvature or a free energy driven collapse is favourable. 
Eventually collapse is seen but only, for this example, when
the drop spans just three posts. Note that for $l=45$ the drop stops moving
once it is fully inside the posts as its free energy becomes independent
of height: it forms a liquid bridge connecting several neighbouring posts.

Indeed, we expect from the two dimensional calculations that collapse 
is preempted by depinning for posts with $b/a \sim 1$ until the drops are 
very small. In three dimensions depinning will be even more important because the 
receding contact angle is larger than $\theta_e$, its value in two dimensions, because 
the distortion of the interface makes it more favourable for the drop to depin. 
From Fig. \ref{figQuasi2D}(a) and (b), we obtain $\theta_R \sim 120^\mathrm{o}$.

\section{Summary}

To conclude, we have investigated the behaviour of an evaporating drop
on a superhydrophobic surface. As the drop volume decreases quasi-statically it can move in
three ways: (i) the drop attains its receding contact angle and the
contact line moves inwards across the surface (ii) the free energy
barrier to collapse vanishes and the drop moves smoothly down the
posts (iii) the drop touches the base of the surface patterning and
immediately collapses. The depinning (i) is predominant and, unless
the posts are widely spaced, or the surface is only very weakly
hydrophobic, collapse occurs only for drops spanning a very small number of
posts.

This suggests strategies that could be used to suppress transitions to
the Wenzel state. Long enough posts are needed to prevent
curvature-driven collapse, i.e. $l \gtrsim b^2/R$, and the free energy barrier to the
transition can be enhanced by choosing $\theta_e$ as large as possible
and using closely spaced posts, i.e. $b \lesssim a$. A mobile contact line will also help
as this will relax any build up of curvature.

Our results are in line with recent experiments \cite{Mchale1,Quere2,Moulinet2}. 
In \cite{Quere2}, for long posts, the contact line retreated as the drop shrank 
and collapsed only at the very end of evaporation. For short posts, a few depinning 
events were followed by collapse at a radius consistent with a curvature-driven 
mechanism, $R_c \propto b^2/l$. It is not clear, however, whether the drop interface 
was suspended on all the posts or just those at the rim at the point of collapse: 
this detail is important in determining the constant of proportionality. In \cite{Moulinet1},
the various drop configurations found here are also observed, including the 
depinning of the drop from all but the outer posts. In 
\cite{Moulinet2}, for the somewhat different situation of drops bounced onto a 
surface, the critical pressure for impalement varied linearly with post height 
for short posts, as expected for curvature-driven collapse, and showed a clear 
crossover to a length-independent regime for longer posts, consistent with a drop 
overcoming a free energy barrier.

\acknowledgments
We thank D. Qu\'{e}r\'{e} for bringing this problem to our attention and for pointing
out the collapse mechanism due to the centre of the interface touching the base surface.
We appreciate useful discussions with G. Alexander, G. McHale and S. Moulinet. HK 
acknowledges support from a Clarendon Bursary and the INFLUS project.

\end{document}